\documentclass[12pt,article]{amsart2000}
\usepackage{amsfonts}
\RequirePackage{amsmath2000}
\newcommand{\cc}{\mathcal{C}}
\newcommand{\ca}{\mathcal{A}}
\newcommand{\cb}{\mathcal{B}}

\newcommand{\ce}{\mathcal{E}}
\newcommand{\cg}{\mathcal{G}}

\newcommand{\cf}{\mathcal{F}}
\newcommand{\cs}{\mathcal{S}}
\newcommand{\al}{{\alpha}}
\newcommand{\be}{{\beta}}

\newcommand{\Ga}{{\mathit\Gamma}}

\newcommand{\si}{{\sigma}}

\newcommand{\ba}{{\mathbb A}}
\newcommand{\bc}{{\mathbb C}}

\newcommand{\fm}{{\mathfrak M}}

\newcommand{\lto}{{\longrightarrow}}
\begin{document}

\title{\bf On localizing topological algebras}

\author{\bf Anastasios Mallios}

\date{}

\maketitle

\pagestyle{myheadings} \markboth{\centerline {\small {\sc
{Anastasios Mallios}}}}
         {\centerline {\small {\sc {On localizing topological algebras}}}}

\begin{quote}{\footnotesize {\bf Abstract.} Through the subsequent discussion we consider a certain particular
sort of (topological) algebras, which may substitute the ``{\em
structure sheaf algebras}'' in many---in point of fact, in
all---the situations of a geometrical character that occur, thus
far, in several mathematical disciplines, as for instance,
differential and/or algebraic geometry, complex (geometric)
analysis etc. It is proved that at the basis of this type of
algebras lies the {\em sheaf-theoretic} notion of (functional)
{\em localization}, which, in the particular case of a given
topological algebra, refers to the respective ``{\em Gel'fand
transform algebra}'' over the spectrum of the initial algebra. As
a result, one further considers the so-called ``{\em geometric
topological algebras}'', having special cohomological properties,
in terms of their ``{\em Gel'fand sheaves}'', being also of a
particular significance for (abstract) differential-geometric
applications; yet, the same class of algebras is still ``{\em
closed}'', with respect to appropriate {\em inductive limits}, a
fact which thus considerably broadens the sort of the topological
algebras involved, hence, as we shall see, their potential
applications as well.}
\end{quote}

\setcounter{section}{-1}
\begin{centering}
\section{Introduction}                              
\end{centering}

Our aim by the present paper is to present a certain particular
type of topological algebras, that seems to be at the basis of
what we may call {\it ``geometric} (topological) {\it algebras''},
in the sense that this sort of (topological) algebras are, indeed,
fundamental and, in point of fact, determine the inner structure
of several important mathematical disciplines of a {\it
geometrical character}, as, for instance, {\it differential
geometry} (of smooth manifolds), {\it geometry of} complex
(analytic) manifolds, through ($\bc$-, or even, vector-valued)
{\it holomorphic functions, algebraic geometry} (commutative
case), and the like. Our study is essentially {\it
sheaf-theoretic}, given that sheaf theory is, as we shall see, the
appropriate set-up to formulate and treat structural properties of
the topological algebras, under consideration. The present account
may also be considered, as a natural continuation and even further
improvement and/or extension of our previous study in [12], [13].

We start, by first giving, in the next Section, the necessary
preliminary material, as well as, the relevant results, concerning
the sheaf-theoretic rudiments for the study of the topological
algebras in the title of this paper and their further
(geometrical) applications.

\begin{centering}
\section{Functional presheaves and their (functional) localizations} 
\end{centering}

Suppose we are given a {\it topological space} $X$, along with a
{\it functional presheaf} on $X$,
\begin{equation}        
    A\equiv \{ A(U):U\subseteq X, \ \textit{open}\},
\end{equation}
in the sense that one has (by definition of the concept at issue),
\begin{equation}        
    A(U)\subseteq \cf (U,Y), \quad \text{for any \it{open}} \quad
    U\subseteq X,
\end{equation}
the second member of (1.2) denoting the {\it set of maps} (not
a\;\;p\;r\;i\;o\;r\;i continuous) on $U\subseteq X$ in the set
(space) $Y$, which in the sequel is identified, for convenience,
with the {\it complex numbers} $\bc$ (the general case being,
however, also of a particular interest; see, for instance, Note
2.1 in the sequel). So $A(U)$ is just a given {\it set of
$\bc$-valued maps} on the {\it open} set $U\subseteq X$, hence,
the previously applied terminology. In this connection, we first
remark that:

\bigskip \noindent (1.3)\hfill
\begin{minipage}{11cm}
    {\it any} given {\it presheaf} (of sets) {\it on $X$ ``is''
    converted}, via its {\it shea\-fi\-fi\-ca\-tion, into a functional
    presheaf on $X$}.
\end{minipage}

\bigskip \noindent
Concerning the terminology on {\it sheaf theory}, applied
herewith, see e.g. A.\;Mallios [14: Vol. I; Chapt. I]. Yet, for
the inverted commas in {\it ``is''}, as above, see also (1.8),
(1.11) in the sequel. Thus, expressing (1.3), otherwise, in the
natural question, whether {\it ``functional presheaves''} are so
important, one can say that, indeed, they are, in point of fact,
{\it the only ones} (!), since,

\bigskip \noindent $(1.3)'$\hfill
\begin{minipage}{11cm}
    {\it any} given {\it presheaf is converted}, along the way of its
    {\it ``sheafification''} (:\;{\it ``completion''}, \`{a} la Leray), {\it into
    a functional one}.
\end{minipage}

\bigskip
Now, according to the general theory (loc. cit.), {\it for any
given presheaf} (of sets, not necessarily functional), {\it one
obtains a sheaf}, its {\it sheafification},
\setcounter{equation}{3}
\begin{equation}        
    \cs (A) \equiv \ca \equiv (\ca, \pi, X), \ \text{viz. a {\it
    map}} \ \pi : \ca \lto X,
\end{equation}
such that one has, by definition,
\begin{equation}        
    \ca := \sum_{x\in X} \ca_x ,
\end{equation}
where we still set;
\begin{equation}        
    \ca_x \equiv \pi^{-1}(x):= A_x := \varinjlim_{U\in \cb(x)} A(U),
\end{equation}
with $U$ in the last ``limit'' varying over a fundamental system
$\cb (x)$ of open neighborhoods of the point $x\in X$. Thus, one
gets at a map (see (1.1))
\begin{equation}        
    \rho \equiv (\rho_U):A\equiv \{ A(U)\} \; \lto \; \Ga (\ca) \equiv \{
    \Ga (U,\ca)\equiv  \ca (U) \} ,
\end{equation}
such that one has;
\begin{equation}        
    \rho_U :A(U) \; \lto \; \rho_U (A(U))\equiv \rho (A(U))\equiv
    \rho(A)(U) \equiv A^\sim (U) \subseteq \ca (U)\equiv \Ga (U,\ca ),
\end{equation}
where, in particular, {\it we set}
\begin{equation}        
    \rho_U (s)\equiv \tilde{s}:U \; \lto \; \ca :x\longmapsto \tilde{s}
    (x)\\
    := \rho^{x}_{U} (s)\in A_x =\ca_x \subseteq \ca ,
\end{equation}
{\it for any} $s\in A(U)$, while
\begin{equation}        
    \rho^{x}_{U}:A(U) \; \lto \; \ca_x =A_x
\end{equation}
is the corresponding to (1.6) {\it canonical map}, for any $U\in
\cb (x)$ (see also ibid., p. 30; (7.9)). Now, a given {\it
presheaf} (of sets) $A$ on $X$, as above, is said to be a {\it
momopresheaf}, if each one of the maps $\rho_U$, as in (1.8), is
{\it one-to-one}, for any {\it open} $U\subseteq X$. It is easy to
see, by the very definitions, that

\bigskip \noindent (1.11) \hfill
\begin{minipage}{11cm}
    the condition of {\it being a} given {\it presheaf a
    monopresheaf}\;\,is {\it equi\-va\-lent with the first} one {\it of
    the two} classical {\it conditions}, \`{a} la Leray, {\it of
    being the} given {\it presheaf ``complete''}, viz. actually a
    {\it sheaf}.
\end{minipage}

\bigskip \noindent
See loc. cit., Chapt. I; p. 46, Section 11, in particular, p.
51ff, proof of Proposition 11.1, yet, p. 54; (11.36$'$), along
with p. 47, Remark 11.1, i).\;\rule{1.3ex}{1.3ex} As an immediate
consequence of the above, one obtains that;

\bigskip \noindent (1.12)\hfill
\begin{minipage}{11cm}
    {\it any functional presheaf} (cf.\;(1.1), (1.2)) {\it is}, in
    effect, {\it a mo\-no\-pre\-sheaf}.
\end{minipage}

\bigskip \noindent
As a result of the preceding, we remark that:

\bigskip \noindent (1.13)\hfill
\begin{minipage}{11cm}
    the {\it sheafification} of a given (abstract) presheaf is
    just a manner of {\it converting} its {\it elements} (:
    elements of the sets $A(U)$, $U\subseteq X$, cf. (1.1)) {\it
    into} ({\it generalized}) {\it functions}, viz. {\it sections}
    of the corresponding {\it sheaf} (:\;{\it complete presheaf}).
\end{minipage}

\bigskip \noindent
In particular, by considering the so-called {\it Gel'fand
presheaf} of a given {\it topological algebra}, thus, by its very
definition, as we shall see right below, a {\it functional
presheaf} over the {\it spectrum} (:\;{\it Gel'fand space}) of the
given topological algebra, the obstruction to the above procedure,
as in (1.13), that is, to {\it converting the elements of the
algebra} at issue {\it into sections} (of an appropriate sheaf,
viz. of the respective {\it Gel'fand sheaf}) is measured exactly
by the non-{\it localness} of the topological algebra, under
consideration, a notion of our main concern, which will be
examined right away, by the next Section.

However, as a preamble to that study, we first turn our attention
to the way one succeeds, in {\it ``completing'' a} given {\it
functional presheaf} (cf. (1.2)) {\it into a ``complete
presheaf''}, viz. into a {\it sheaf}, thus, alias, we are looking,
first, at the so-called {\it ``functional completion''} of the
initial (functional) presheaf. As we shall see, right in the
sequel, {\it this procedure consists} actually {\it in}

\bigskip \noindent (1.14)\hfill
\begin{minipage}{11cm}
    {\it adding} to the given (functional) presheaf {\it all
    locally defined functions}, which further {\it locally belong},
    as well, {\it to the} initial {\it presheaf}.
\end{minipage}

\bigskip \noindent
In that sense, we can already understand, in anticipation (see
also (1.21) below), the reason that some important classical {\it
functional presheaves}, as, for instance, those of {\it
continuous, differentiable, holomorphic} functions, are, in
effect, {\it sheaves}, that is, {\it complete presheaves}\,; thus,
they cannot be further completed, their elements being by their
very definition, functions, {\it locally characterizable}, hence,
already elements of the given presheaves. In other words, we can
further state, in anticipation (see thus Section 2 in the sequel),
that:

\bigskip \noindent (1.15)\hfill
\begin{minipage}{11cm}
    a {\it functional presheaf}, consisting of elements, that
    can be {\it locally characterized}, is virtually a {\it complete
    presheaf}\,; hence, a {\it sheaf} (viz., in effect, isomorphic
    to its {\it sheafification}, that is, to the sheaf, it
    generates, cf. Section 2 below). That is, {\it a functional
    presheaf, as before, coincides}, in effect, {\it with its
    ``localization''} being thus a {\it complete presheaf}, hence
    (Leray), a {\it sheaf} (see also (1.21) in the sequel,
    along with subsequent comments therein).
\end{minipage}

\bigskip
So we first explain, more precisely, the terminology we employed
in (1.14): That is, by considering a {\it functional presheaf}, as
in (1.1), (1.2), {\it ``its elements''}, viz. elements of the
particular sets $A(U)$, with $U$ open in $X$, are, by definition,
numerical-valued (for convenience, $\bc$-{\it valued}) {\it
functions, ``locally defined''} on the various {\it open} sets
$U\subseteq X$. Yet, we also set the following crucial definition.
That is, one has.

\bigskip                                    
{\bf Definition 1.1.} Suppose we have a {\it functional presheaf}
\setcounter{equation}{15}
\begin{equation}        
    A\equiv \{ A(U):U \ \textit{open in} \ X\}
\end{equation}
(see (1.1), (1.2)). Then, a given $\bc$-{\it valued map} $\al
:U\to \bc$, with $U$ {\it open} in $X$, {\it locally belongs to}
$A$, if, for every point $x\in U$, there is an open set
$V\subseteq X$, with $x\in V\subseteq U$, and an element (:\;{\it
``local function''}) $h\in A(V)$, such that;
\begin{equation}        
    \al =h\big|_V .
\end{equation}
(:\;precisely, $\al \big|_V =h$).

\medskip
Now, the previous definition leads naturally to the important
notion for our purpose of the {\it localization of a} given
(functional) {\it presheaf}\,: That is, suppose we are given a
{\it functional presheaf} $A$, as in (1.16) (see also (1.2)).
Then, by {\it definition},

\bigskip \noindent (1.18) \hfill
\begin{minipage}{11cm}
    the {\it localization of} $A$, say $\tilde{A}$, consists of
    those, locally defined $\bc$-va\-lu\-ed functions on $X$, which
    locally belongs to $A$ (cf. Definition 1.1).
\end{minipage}

\bigskip \noindent
Thus, technically speaking, given $A$, as above, one considers its
{\it localization} $\tilde{A}$, viz. the presheaf on $X$, given by
\setcounter{equation}{18}
\begin{equation}        
    \tilde{A} \equiv \{ (\tilde{A}(U):U \ \text{\it{open} in}
    \ X); \ \tilde{\rho}^{\,U}_{\,V} \} ,
\end{equation}
where one sets (Definition 1.1);
\begin{equation}        
\begin{aligned}
    \tilde{A}(U):=\{ & \al :U \lto \bc | \ \forall \ \, x\in U \ \exists \ \, V
    \ \ \text{{\it open} in} \ \ X, \ \ \text{with}\\
    & x\in V\subseteq U, \ \ \text{and} \ \ h\in A(V), \ \
    \text{such that} \ \ \al \big|_V =h \} .
\end{aligned}
\end{equation}
Of course, $\tilde{A}$ is a {\it presheaf} on $X$; moreover, by
its very definition, it is a {\it functional presheaf} (cf.
(1.20)), hence, a {\it monopresheaf} (see (1.12)). Indeed, our
claim is that, in effect, $\tilde{A}$ {\it is} {\it a complete
presheaf} on $X$ (Leray's definition): That is, given an open
covering $(U_i)_{i\in I}$ of an {\it open} $U\subseteq X$, viz.
$U=\displaystyle{\bigcup_{i}}U_i$, and an element $(\al_i )\in
\displaystyle{\prod_{i}}\tilde{A}(U_i)$, with $\al_i =\al_j
\big|_{U_{ij}\equiv U_{i}\cap U_{j}\not= \emptyset}$, {\it one has
to prove that} these exists $\al \in \tilde{A}(U)$, with $\al
\big|_{U_i}=\al_i$, $i \in I$; indeed, the proof follows
straightforwardly from Definition 1.1 and (1.20), along with our
hypothesis for the family $(\al_i)$.\;\rule{1.3ex}{1.3ex} Thus, we
have proved, so far, the basic result that:

\bigskip \noindent (1.21)\hfill
\begin{minipage}{11cm}
    {\it the localization of a} given {\it functional presheaf}
    (as defined by (1.20)) {\it is a complete presheaf}\,; hence
    (Leray's definition), a {\it sheaf}.
\end{minipage}

\bigskip \noindent
See also, for instance, A.\;Mallios [14: Chapt. I; Section 11],
concerning the terminology applied in (1.21). Thus, in view of the
preceding, we fully understand now our previous comments in
(1.15), since {\it a functional presheaf, whose elements} (:\;{\it
local functions}) {\it are locally characterized, coincides}, in
point of fact, {\it with its localization}, according to the very
definition of the sets in (1.20) and the term {\it ``locally
characterized'', (1.20) being}, in effect, {\it the precise
definition of the latter term}\,(!). In other words,, in view of
(1.21), {\it the} given {\it presheaf is complete}.

We turn now, by the next Section 2, to describe (1.19), in terms
of {\it ``germs''} {\it of functions}, equivalently, to look at
the notion of a sheaf, as a {\it ``local homeomorphism''}
(concerning {\it the map $\pi$, as in (1.4)}; Lazard's
definition), and further connecting it with our previous
discussion. For technical details, we still refer to A.\;Mallios
[14: Chapt. I; Sections 7, 8].

\bigskip
\begin{centering}
\section{Sheafification of a functional presheaf}   
\end{centering}

\begin{flushright}
\begin{minipage}{7cm}
    {\it ``...\;to understand what's what ...\;is a vital
    aspect of Mathematics''}. 
\begin{flushright}
\begin{minipage}{6,5cm}
    S.\;Mac Lane, in {\it ``Mathematics Forms
    and Function''} (Springer-Verlag (1986), p. 288).
\end{minipage}
\end{flushright}
\end{minipage}
\end{flushright}

By considering a {\it functional presheaf} $A$ on a topological
space $X$, as above (see (1.1), (1.2)), and then its {\it
sheafification} (cf. (1.5), (1.6)), \setcounter{equation}{0}
\begin{equation}        
    \cs (A)\equiv \ca ,
\end{equation}
our main objective in this Section is to prove the relation
(:\;{\it isomorphism of sheaves})
\begin{equation}        
    \cs (A)\equiv \ca \cong \tilde{A} ;
\end{equation}
in other words, we claim that:

\bigskip \noindent (2.3)\hfill
\begin{minipage}{11cm}
    if we {\it fill up a} given {\it functional presheaf} $A$ on
    a topological space $X$, {\it by} all those ({\it local}) {\it
    functions}, that {\it locally belong to it} (Definition 1.1)
    (viz., by looking at the {\it localization $\tilde{A}$ of}
    $A$, cf. (1.19), (1.20)), then {\it the resulting} ({\it
    functional}) {\it presheaf $\tilde{A}$ is} ({\it complete},
    cf. (1.21), and, in point of fact, {\it isomorphic to}) {\it
    the sheafification of the} given {\it presheaf} $A$.
\end{minipage}

\bigskip \noindent
Consequently,

\bigskip \noindent (2.4)\hfill
\begin{minipage}{11cm}
    the {\it ``completion''} of a given functional presheaf $A$, to
    become a complete presheaf, hence, to its {\it sheafification,
    ``is''} virtually reduced to {\it its localization} $\tilde{A}$. We also
    speak of $\tilde{A}$, by extending herewith a relevant
    terminology of R.\;Arens [1], as the {\it hull of $A$, with
    respect to $X$}.
\end{minipage}

\bigskip \noindent
Thus, in other words,

\bigskip \noindent (2.5)\hfill
\begin{minipage}{11cm}
    a given {\it functional presheaf} on a topological space $X$
    is {\it complete}, viz. it coincides with the (complete
    presheaf of sections of the) sheaf it generates, if, and only
    if, it is also {\it ``functionally complete''}; that is,
    equivalently, if, and only if, it {\it contains any} ({\it
    local}) {\it function that locally belongs to it}.
\end{minipage}

\bigskip \noindent
Indeed, the claimed {\it isomorphism} (2.2) is actually referred
to an {\it isomorphism of the complete presheaves} concerned; that
is, {\it the complete presheaf of local sections of $\ca \equiv
\cs (A)$ is isomorphic to} $\tilde{A}$, the latter being also a
{\it complete presheaf} on $X$, cf. (1.21)). Namely, one has to
prove the following (set-theoretic) {\it bijection};
\setcounter{equation}{5}
\begin{equation}        
    \tilde{A}(U)=\Ga (U,\ca )\equiv \ca (U),
\end{equation}
{\it for any open} $U\subseteq X$: Thus, for any $\al \in
\tilde{A}(U)$, one defines a map
\begin{equation}        
    \ca(U)\ni \tilde{\al}:U\lto \ca :x\longmapsto \tilde{\al}(x):=
    [s]_x \equiv \tilde{s}(x) \in \ca_x \subseteq \ca
\end{equation}
(see also (1.5)), where, by definition of $\al$ (cf. (1.20)), one
has
\begin{equation}        
    \al \big|_V =s\in A(V)\underset{\,\to}{\subset} \ca (V), \ \
    \textit{with} \ \ x\in V\subseteq U,
\end{equation}
given $x$ in $U$ (see also (1.12)); so (2.7) is well-defined,
hence, also the correspondence
\begin{equation}        
    \tilde{A}(U) \xrightarrow{\;\;i_U \;\;}{}\ca (U):\al
    \longmapsto \tilde{\al} ,
\end{equation}
as given by (2.7). Now, {\it the} same {\it map} $i_U$ {\it is
one-to-one}; this follows straightforwardly, in view of (2.7) and
(2.8). Finally, $i_U$ {\it is onto}: That is, given $s\in \ca
(U)$, one defines a map $t \equiv (t_x )\in \tilde{A} (U)$,
according to the relation;
\begin{equation}        
    t(x):=t_x (x), \ \ x\in V_x \subseteq U=\bigcup_{x\in U} V_x ,
\end{equation}
such that,
\begin{equation}        
    s(x)= \rho^{x}_{V_x}(t_x)\in \ca_x = \varinjlim_{V\in {\mathcal V}(x)}
    A(V),
\end{equation}
{\it with} $t_x \in A(V_x) \underset{\to}{\subset} \ca (V_x)$,
while, by virtue of (2.11), one still obtains;
\begin{equation}        
    t_x =t_y \big|_{V_{xy}\equiv V_x \cap V_y \not= \emptyset},
\end{equation}
which also yields, in view of (2.10) and (1.21), that $t\in
\tilde{A}(U)$. Furthermore, one also gets at the relation,
\begin{equation}        
    \tilde{t}=s\in \ca (U),
\end{equation}
based on (2.7), (2.8), and the preceding last three relations,
which proves the assertion, hence, finally (2.6), as
well.\;\rule{1.3ex}{1.3ex} In sum, one thus obtains the following
{\it isomorphism of complete presheaves} (see also (2.1)),
\begin{equation}        
    \tilde{A}\equiv (\tilde{A}(U), \tilde{\rho}^{U}_{V})\cong \Ga
    (\cs (A))\equiv \Ga (\ca)\equiv (\ca (U),\si^{U}_{V}),
\end{equation}
that also explains the slight {\it abuse of notation} employed,
for convenience, {\it in} (2.2).

\bigskip \noindent $ $\hfill
\begin{minipage}{11cm}
    \hspace{0.5cm}{\bf Note 2.1.}\;$-$ The previous procedure of
    constructing the {\it sheafification} of a given {\it
    functional presheaf}, by just considering its {\it
    localization}, viz. the {\it complete presheaf}, containing
    the initial one, as explained in (2.3), {\it represents}, in
    point of fact, {\it the general case}, as well; viz. the {\it
    sheafification of an} arbirtary (not necessarily functional)
    {\it presheaf}\,: Indeed, this is explained by considering the
    map (1.7), converting a given presheaf $A$ into a {\it
    functional} one $\rho (A)$, whose elements are ({\it local})
    {\it sections} of $\ca$, thus {\it local $\ca$-valued
    functions on} $X$ (cf. (1.9)), $\ca$ being here the final {\it
    sheaf generated} by $A$, and, in effect, {\it by} $\rho (A)$
    (see A.\;Mallios [14:\;Chapt.\;I, p.\;31; (7.12)]); the\-re\-fo\-re,
    $\ca$ (when also
    identified with the {\it complete presheaf of} its
    {\it sections}, $\Ga (\ca)$, cf. (1.7)) {\it is}
    virtually {\it isomorphic}, according to the preceding, with
    the {\it localization} of $\rho (A)$, $\widetilde{\rho(A)}$
    (: localization of the {\it ``functionalization''}, $\rho
    (A)$, of $A$). That is, one actually obtains;
    \begin{equation}    
        \widetilde{\rho (A)}=\Ga (\ca)\equiv \Ga (\cs (A))= \Ga
        (\cs (\rho (A))),
    \end{equation}
    within an {\it isomorphism of complete presheaves} (see
    (1.20), (1.21)). In this connection, cf.\;also loc.\;cit.,
    Chapt.\;I; Sections 3, 7, in par\-ti\-cu\-lar, p. 31, Scholium 7.1.
\end{minipage}

\bigskip \noindent So the moral here is that;

\bigskip \noindent (2.16)\hfill
\begin{minipage}{11cm}
    {\it sheafifing a given presheaf} (\`{a} la Leray), {\it
    means}, in effect, {\it localize its ``functionalization''}
    (: the presheaf obtained, by converting the elements of the
    initial presheaf into functions (indeed, {\it sections}), cf.
    (1.9)).
\end{minipage}

\bigskip \noindent Yet, roughly speaking, we may still say that;

\bigskip \noindent (2.17) \hfill
\begin{minipage}{11cm}
    {\it ``sheafifing''} means, in point of fact, {\it
    ``localize''}.
\end{minipage}

\bigskip By closing the preceding discussion, we still remark, in
passing, that our previous considerations, pertaining to the {\it
sheafification} of a given {\it functional presheaf}, through its
{\it localization}, apart from its usual application in important
classical examples, as already mentioned in the foregoing (see
also loc. cit., Chapt. I; p. 17ff, Section 4), one also encounters
another interesting justification of the same point of view in
some recent {\it applications} of the notion {\it of sheaf} in
{\it Quantum Field Theory}, or even in {\it Quantum Relativity};
see thus, for instance, R. Haag [7: p. 326], I. Raptis [26],
A.\;Mallios-I.\;Raptis [19]. To quote e.g. R.\;Haag (ibid.), one
realizes that (italicization below is ours);

\bigskip \noindent (2.18) \hfill
\begin{minipage}{11cm}
    {\it ``... the central message of Quantum Field Theory} [is]
    {\it that all information characterizing the theory is
    strictly local ...''}
\end{minipage}

\bigskip \noindent
Now, based on the preceding, we also note that the above {\it
``local information''} is further improved, through the {\it
``functional completion''}, as before, that is, by {\it adding}
thus to any given initial (local) information {\it all the}
possible {\it ``equivalent'' ones}, a function (:\;procedure),
which, in point of fact, lies at the same basis of implementing
(cf. (1.20)) the very notion of {\it sheaf} in our arguments (see
also (2.3) and/or (2.15)). Accordingly, the sort of application of
the foregoing in nowadays quantum field theory, as alluded to
above. Yet, we can still conclude, in view of the foregoing (cf.,
in particular, (2.15), as above), that, in other words,

\bigskip \noindent (2.19) \hfill
\begin{minipage}{11cm}
    by considering the {\it sheaf}, generated  by a given {\it
    presheaf}, means, in point of fact, to

    \medskip \noindent (2.19.1)\hfill
    \begin{minipage}{9cm}
        {\it collect together all}
        the {\it local information}, we can get, on the basis of that
        one, which is, already, {\it locally afforded, by the elements
        of the} given {\it presheaf}.
    \end{minipage}
\end{minipage}

\bigskip \noindent
The above still supports the aforementioned applicability of the
notion of {\it sheaf} in {\it quantum theory} and/or {\it quantum
relativity}; see also, for instance, H.F.\;de Groote [6].

\newpage
\begin{centering}
\section{Localization of topological algebras}      
\end{centering}

\begin{flushright}
\begin{minipage}{7cm}
    {\it ``There are many advantages in developing a theory in the most
    general context possible.''}

\begin{flushright}
\begin{minipage}{6cm}
    R.\;Hartshorne, in {\it ``Algebraic Geometry''}
    (Springer-Verlag, New York, 1977). p. 59.
\end{minipage}
\end{flushright}
\end{minipage}
\end{flushright}

As the title of this Section indicates, our main objective by the
ensuing discussion is to apply our previous considerations in the
particular case, when one has to deal with {\it $($pre$)$sheaves
of topological algebras}, in the general framework of which, one
can still examine, as already hinted at in the preceding, the sort
of topological algebras, we are looking at, by the present study.

We start, by presenting the relevant general set-up: Thus, first,
by a {\it topological algebra space}, we mean a pair,
\[
    (\ba ,\;\fm (\ba )),
    \leqno{(3.1)}
\]
consisting of a given {\it topological algebra} $\ba$, together
with its (global) {\it spectrum} (alias, {\it Gel'fand space}),
$\fm (\ba)$; of course, we still posit here, by definition, that
the latter (a {\it Hausdorff completely regular}) space is not
trivial (:\;empty). Yet, we refer to A. Mallios [11] for the
relevant terminology employed herewith.

Now, a topological function algebra, which is naturally associated
with a given topological algebra space, as in (3.1), is the
algebra
\[
    \cc_c (\fm (\ba)),
    \leqno{(3.2)}
\]
that is, the set of $\bc$-valued continuous functions on $\fm
(\ba)$, being a $\bc$-algebra, with point-wise defined operations,
further endowed with the compact-open topology, becoming thus, in
particular, a {\it locally $m$-convex} (topological) {\it algebra}
(ibid.). Yet, an important {\it representation} of $\ba$ into the
latter algebra is achieved, via the so-called {\it Gel'fand
representation}, and the associated with it {\it Gel'fand map},
\[
    \cg_\ba \equiv \cg :\ba \; \lto \; \cc (\fm (\ba )),
    \leqno{(3.3)}
\]
such that,
\[
    \cg_\ba (x)\equiv \hat{x}:\fm (\ba) \; \lto \; \bc :f \longmapsto
    \hat{x} (f):=f(x),
    \leqno{(3.4)}
\]
for every $f\in \fm (\ba)$ (loc.\;cit.; we call $\hat{x}$, the
{\it Gel'fand transform of} $x\in \ba$). Thus, we are now in the
position to turn ourselves to our main objective, viz. to look at
the possibility of being a given topological algebra {\it
localizable}, in the sense in which the latter term will become
clear by the subsequent discussion:

So given a {\it topological algebra space}, as in (3.1), one can
naturally associate with it, by means of the Gel'fand
representation, as above, a {\it functional $(\bc$-$)$algebra
presheaf} on $\fm (\ba)$ (:\;{\it base space} of (3.1)), which we
call the {\it Gel'fand presheaf} of the given topological algebra
space, or, for short, just, {\it of} $\ba$, defined by,
\[
    \cg \textit{presh}(\ba )\equiv \hat{\ba}\equiv \{ \ba (U)^{\wedge} :
    U\subseteq \fm (\ba ), \ \textit{open}\},
    \leqno{(3.5)}
\]
where {\it we set} (cf.\;also (3.4), while {\it we} still {\it
put} $\ba^{\wedge}\equiv im \cg$, called the {\it Gel'fand
transform algebra} of $\ba$),
\[
    \ba (U)^{\wedge}:=\ba^{\wedge}\big|_U =\{\hat{x}\big|_U : x\in\ba
    \},
    \leqno{(3.6)}
\]
for every {\it open} $U\subseteq \fm (\ba)$, along with the
corresponding obvious {\it restriction maps},
\[
    \rho^{U}_{V}:\ba^{\wedge}\big|_U \lto \ba^{\wedge}\big|_V ,
    \leqno{(3.7)}
\]
for any {\it open} $U,V$ in $\fm (\ba)$, with $V\subseteq U$.
Therefore, by its very definition,

\bigskip \noindent
(3.8) \hfill \begin{minipage}{11cm}
    {\it the Gel'fand presheaf of} $\ba$ (cf. (3.5)) {\it is a
    functional presheaf} on $\fm (\ba)$, hence (see (1.12)), a
    {\it monopresheaf}.
\end{minipage}

\bigskip \noindent
On the other hand, we are actually interested in finding, what we
may call the {\it Gel'fand sheaf of} $\ba$, denoted in the sequel
by
\[
    \ca ,
    \leqno{(3.9)}
\]
thus, by its very definition, as we shall presently see, a {\it
sheaf of} $(\bc$-$)${\it algebras over} $\fm (\ba)$; yet, we still
refer to $\ca$, as the {\it sheafification} of the given
topological algebra $\ba$.

Thus, by definition, $\ca$, viz.\;the Gel'fand sheaf of $\ba$, is
the sheaf on $\fm (\ba)$, ge\-ne\-ra\-ted by the Gel'fand pesheaf
$\hat{\ba}$ of $\ba$ (cf. (3.5)). That is, we set;
\[          
    \ca := \cs (\hat{\ba}).
    \leqno{(3.9)'}
\]
Therefore, in view of (3.8), and of what has been said in Section
2 (see e.g. (2.3), (2.4) or even (2.17)), one actually obtains the
following.

\bigskip {\bf Theorem 3.1.} Given a {\it topological algebra
space} $(\ba, \fm (\ba))$ (see (3.1)), one has;
\[
    \ca =\tilde{\ba} (\equiv \tilde{\hat{\ba}}),
    \leqno{(3.10)}
\]
within an {\it isomorphism of} (complete pre){\it sheaves}, where
the second member of (3.10) stands for the {\it localization of
the Gel'fand presheaf of} $\ba$. That is, in other words,

\bigskip \noindent (3.11) \hfill
\begin{minipage}{11cm}
    the {\it sheaf of germs of the Gel'fand transforms of the
    elements of} $\ba$ (or else, the {\it Gel'fand sheaf $\ca$ of} $\ba$,
    {\it is} (isomorphic to) {\it the sheaf} (:\;complete
    presheaf) $\tilde{\ba}$ {\it of germs of $\bc$-valued
    functions, locally belonging to
    $\ba^{\wedge}$}.\;\rule{1.3ex}{1.3ex}
\end{minipage}

\bigskip For convenience, we only recall the definition of the
({\it complete}) {\it presheaf} (see (1.21) $\tilde{\ba}$, based
on (1.20); viz. one has, for any {\it open} $U\subseteq \fm
(\ba)$:

\bigskip \noindent (3.12)\hfill
\begin{minipage}{11cm}
    $\tilde{\ba}(U)=\{ \al :U \to \bc | \ \forall \ f\in U \ \exists
    \ V \ \text{{\it open} in} \ \fm (\ba)$,
    with $f\in V\subseteq U$, and $\hat{x}\big|_V \in \ba
    (V)^{\wedge}$ (hence, actually an element $x\in \ba$), such
    that; $\al \big|_V =\hat{x}\big|_V$ (for short, $\al =
    \hat{x}\big|_V )\}$.
\end{minipage}

\bigskip \noindent
As a result, one thus infers, according to the very definitions,
that;

\bigskip \noindent (3.13)\hfill
\begin{minipage}{11cm}
    the {\it localization of the Gel'fand presheaf} of a given
    topological algebra $\ba$ consists of those $\bc$-{\it valued
    local functions on} $\fm (\ba)$, which are {\it locally
    Gel'fand transforms of elements of} $\ba$.
\end{minipage}

\bigskip \noindent
Of course, in view of (3.10), {\it the same characterization}, as
above, {\it is} still {\it valid for the Gel'fand sheaf} itself
{\it of} $\ba$ (when applying {\it Leray's terminology}).

On the other hand, in view of (3.8), {\it one has};
\[
    \hat{\ba}\underset{\;\to}{\subset} \tilde{\ba}
    \leqno{(3.14)}
\]
(within an {\it isomorphism into} of the presheaves concerned; see
also A.\;Mallios [13: Chapt. I; Section 6]). Indeed, the previous
relation leads us now to the following basic.

\bigskip {\bf Definition 3.1.} Given a topological algebra space,
as in (3.1), we say that the topological algebra $\ba$ is {\it
localizable}, whenever one has
\[
    \hat{\ba} =\tilde{\ba},
    \leqno{(3.15)}
\]
viz. whenever the Gel'fand presheaf of $\ba$ (cf. (3.5)) is
already localized, hence (cf. (1.21)), complete.

\medskip Therefore, based on (3.10), in the case of a {\it
localizable topological algebra}, one gets at the following
relation;
\[
    \hat{\ba} = \tilde{\ba} =\ca
    \leqno{(3.16)}
\]
(in the sense of (3.10), concerning the last equality). Thus, as a
straightforward application of our terminology in the preceding,
one still obtains, in the case of a {\it localizable} {\it
topological algebra} $\ba$ (cf. (3.6)),
\[
\begin{aligned}
    \cg (\ba)\equiv \ba^{\wedge}&=\ba^{\wedge}\big|_{\fm (\ba)}=
    \ba (\fm (\ba ))^{\wedge} \equiv \hat{\ba}(\fm (\ba ))\\
    &=\tilde{\ba}(\fm (\ba ))=\Ga (\fm (\ba),\ca )=\ca (\fm (\ba)),
\end{aligned}
    \leqno{(3.17)}
\]
that is, in short, one has, the following {\it isomorphism of
$\bc$-algebras};
\[
    \ba^{\wedge}=\Ga (\fm (\ba),\ca )\equiv \ca (\fm (\ba )),
    \leqno{(3.18)}
\]
viz., in the case of a {\it localizable $($topological$)$ algebra}
$\ba$, the {\it Gel'fand transform algebra of} $\ba$ may be
construed, as the {\it global section algebra of its Gel'fand
sheaf} over $\fm (\ba)$, so that one also speaks then of a {\it
sectional representation of} (the localizable algebra) $\ba$.
However, one actually infers, in full generality that;

\bigskip \noindent (3.19) \hfill
\begin{minipage}{11cm}
    given a {\it topological algebra space} $(\ba ,\fm (\ba ))$,
    then, $\ba$ always {\it admits}, through its Gel'fand transform
    algebra, {\it a sectional representation, via the}
    corresponding {\it Gel'fand sheaf} over $\fm (\ba)$.
\end{minipage}

\bigskip \noindent
Indeed, {\it one has}, in view of (3.10), (3.14) and (3.17), {\it
the relations};
\[              
    \ba^\wedge = \ba (\fm (\ba ))^\wedge \equiv \hat{\ba}(\fm
    (\ba)) \underset{\;\to}{\subset} \tilde{\ba} (\fm (\ba))\cong
    \Ga (\fm (\ba),\ca )\equiv \ca (\fm (\ba)).
    \leqno{(3.20)}
\]
That is, {\it the Gel'fand transforms of the elements of} $\ba$,
can always be viewed as {\it global} ({\it continuous}) {\it
sections over $\fm (\ba)$ of the Gel'fand} sheaf of $\ba$. (Thus,
one gets at a more {\it intrinsic} (direct) {\it sectional
representation} of $\ba$, than that one obtained, anyway, through
the Gel'fand tranform algebra $\ba^\wedge$, as continuous
$\bc$-valued functions on $\fm (\ba)$, cf. (3.3)).

Now, it may happen that the {\it Gel'fand map} of $\ba$, as given
by (3.4), is {\it one-to-one}; equivalently, the points of\;$\fm
(\ba)$ (:\;continuous characters of $\ba$) separate the elements
of $\ba$. We call then $\ba$, a (functionally) {\it semisimple}
topological algebra.

Thus, a topological algebra $\ba$, as in (3.1), which is {\it
localizable} and also {\it semisimple}, is just called a {\it
local topological algebra}. Hence, in that case, one obtains, in
view of (3.18), the relation;
\[              
    \ba \underset{\cg}{\cong} \ba^\wedge =\ca (\fm (\ba))\equiv \Ga
    (\fm (\ba),\ca),
    \leqno{(3.21)}
\]
within {\it isomorphism of $\bc$-algebras}. [{\it Caution}\,! The
term {\it ``local''}, applied herewith, is to be distinguished
from a similar one used in Algebra, pertaining to {\it algebras
with just one maximal} (2-sided) {\it ideal}\,].

The previous notion lies, indeed, at the basis of what we consider
further, as a {\it geometric topological algebra}, a notion that
mostly occurs in several applications of a geometrical character;
see e.g. our previous study in A.\;Mallios [11; 13]. However,
before we proceed to that aspect, in the next Section 4, we first
comment a bit more on some {\it particular instances of local
topological algebras}:

Thus, it is well-known that {\it not every topological algebra is
localizable}; this is simply also the case, even for a unital
commutative Banach algebra. Indeed, one has here the well-known
{\it Eva Kallin's counter example} [10]. On the other hand, about
the same time R.\;Blumenthal [2; 3], based on earlier work of
S.J.\;Sidney [27; 28], deciphered Kalin's example, by supplying an
{\it abstract method of constructing non-local function} ({\it
Banach}) {\it algebras}. Quite recently, R.I.\;Hadjigeorgiou [8;
9] was able to extend the relevant part of the aforementioned work
of Sidney and Blumenthal to the general context of {\it
topological algebra theory}, providing thus, in turn, a
corresponding {\it machinery of constructing}, \`{a} la
Blumenthal, {\it non-local topological} (non-normed) {\it
algebras} [9]. Yet, lately, A.\;Oukhouya proved in his Thesis [24;
25] that {\it every regular locally $m$-convex uniform algebra is
localizable}, supplying thus another aspect of a {\it ``local
theorem''} for topological algebras, yet, with {\it non compact
spectra}, that was already the case in [12: p. 307, Lemma 2.1]. In
this connection, see also previous relevant work of R.M.\;Brooks
[5], where he employs {\it ``partitions of unity''}.

\begin{centering}
\section{Geometric topological algebras}        
\end{centering}

The topological algebras, referred to in the title of this
Section, appear, as, of course, their very name indicates, in
particular geometrical contexts, that we have already hinted at in
the preceding, including, among others, the {\it ``structure
algebras''} (another synonym of ``geometric'') of (geometric)
complex (analytic) function theory, and, in particular, in
differential geometry, this latter case being also our main
motivation for the ensuing discussion.

Thus, for any given {\it topological algebra space}
\[              
    (\ba, \,\fm (\ba)),
    \leqno{(4.1)}
\]
as in (3.1), and under suitable conditions for $\ba$, one is able
to use $\ba$, or rather its corresponding {\it Gel'fand sheaf}
$\ca$ on $\fm (\ba)$ (cf. Theorem 3.1), as the {\it ``sheaf of
coefficients''}, for an extended {\it ``differential-geometric''}
context on $\fm (\ba)$. This type of applications has already
considered in A.\;Mallios [14: Chapt. XI], along with further
potential applications. Now, the aforementioned type of
applications, led us here to associate with the sort of
topological algebras, alluded to by the title of this Section, a
more restricted version of the same notion, than that one,
previously applied in A.\;Mallios [12]. Namely, we set the
following.

\bigskip {\bf Definition 4.1.} Given a topological algebra space
(see (4.1)), we say that $\ba$ is a {\it geometric topological
algebra}, whenever one has;
\[              
    \ba = \Ga (X,\ca)=H^0 (X,\ca),
    \leqno{(4.2)}
\]
within a $\bc$-{\it algebra isomorphism} of the algebras
concerned, while we also assume that,
\[              
    H^p (X,\ca)=0, \ \ p\ge 1,
    \leqno{(4.3)}
\]
where $\ca$ is a $\bc$-{\it algebra sheaf} on $X$, the latter
being a {\it topological space} of the same {\it homotopy type,
as}\;\,$\fm (\ba)$.

\bigskip
The {\it ``cohomology groups''}, in point of fact, $\ca (X)$-{\it
modules} appeared in the last relations above, are meant in the
sense of {\it sheaf cohomology} theory; see, for instance,
A.\;Mallios [14: Chapt. III; yet, in particular, p. 234, Lemma
8.1]. Indeed, the ``geometric topological algebras'' in the sense
of our previous work in [12] (see also [14: Chapt. XI]), are still
such, in the point of view of the above Definition 4.1; thus, cf.
[14: Chapt. XI; p. 320, (4.19), as well as, Chapt. III; p. 238,
(8.24)]). On the other hand, we further remark that the sort of
topological algebras that might be geometric, in the sense of the
above Definition 4.1, can be associated, in effect, with a larger
spectrum of topological algebras, than that one, we might suspect,
at first sight, just, by virtue of the following, indeed, quite
general result. That is, one gets at the next.

\bigskip {\bf Theorem 4.1.} Let
\[              
    (E_\al ,f_{\be \al})
    \leqno{(4.4)}
\]
be an {\it inductive system of unital geometric topological
algebras} (Definition 4.1), having {\it compact spectra}, and let
\[              
    E=\varinjlim E_\al
    \leqno{(4.5)}
\]
be the corresponding {\it inductive limit topological algebra}
(see A.\;Mallios [11: p. 115, Lemma 2.2]. Then, the {\it
topological algebra space}
\[              
    (E,\,\fm (E)=\varprojlim \fm (E_\al ))
    \leqno{(4.6)}
\]
(see also loc. cit., p. 156, Theorem 3.1), defines $E$, as a
(unital) {\it geometric topological algebras}, as well.

\bigskip {\bf Proof.} As a result of topological algebra theory
(loc. cit.), one obtains, based on our hypothesis, the {\it
topological algebra space (4.6)}, having $\fm (E)$, a {\it compact
Hausdorff space}, such that one has;
\[              
    \fm (E)=\varprojlim \fm (E_\al),
    \leqno{(4.7)}
\]
within a {\it homeomorphism of} the {\it topological spaces}
concerned. Thus, by considering the canonical maps,
\[              
    ^{t}f_\al \equiv \phi_\al :\fm (E)\lto \fm (E_\al), \ \ \al
    \in I,
    \leqno{(4.8)}
\]
(ibid., pp. 152, 156; (3.2), (3.25)), associated, by our
hypothesis, with the corresponding herewith, {\it topological
algebra spaces},
\[              
    (E_\al , \,\fm (E_\al )), \ \ \al \in I,
    \leqno{(4.9)}
\]
one further looks at the resulting $\bc$-{\it algebra sheaf} on
$\fm (E)$,
\[              
    \ce := \varinjlim \phi^{*}_{\al} (\ce_\al),
    \leqno{(4.10)}
\]
{\it inductive limit} of the {\it pull-backs} on $\fm (E)$ of the
corresponding to (4.9) $\bc$-{\it algebra sheaves} $\ce_\al$ on
$\fm (E_\al)$, $\al \in I$. (For the terminology applied herewith,
see also, for instance, A.\;Mallios [14: Chapt. I; p. 79ff,
Subsection 14.(b)]). Now, our claim is that;

\bigskip \noindent (4.11)\hfill
\begin{minipage}{11cm}
    {\it the $\bc$-algebra sheaf $\ce$ on} $\fm (E)$, as given by
    (4.10), {\it defines}
    \[          
        (E=\varinjlim E_\al , \ \fm (E)=\varprojlim \fm (E_\al))
        \leqno{(4.11.1)}
    \]
    (see also (4.5), (4.7)), {\it as a geometric topological
    algebra}.
\end{minipage}

\bigskip \noindent
Indeed, the assertion is a straightforward application of the
standard {\it ``continuity theorem'' for $($sheaf$)$ cohomology}
on compact (Hausdorff) spaces: see, for instance, G.E. Bredon [4:
p. 102, Theorem 14.4], along with A.\;Mallios [14: Chapt. III; p.
173ff, Section 4]. That is, {\it one has};
\[              
\begin{aligned}
    H^p (\fm (E),\,\ce )&=H^p (\varprojlim \fm (E_\al), \, \varinjlim
    \phi^{*}_{\al}(\ce_\al ))\\
    &=\varinjlim H^p (\fm (E_\al),\,\ce_\al), \ \ p\geq 0,
\end{aligned}
    \leqno{(4.12)}
\]
that yields the assertion, according to our hypothesis; cf. also
(4.1) and (4.2).\;\rule{1.3ex}{1.3ex}

\bigskip $ $\hfill
\begin{minipage}{11cm}
    \hspace{0.5cm}{\bf Note 4.1.}$-$ The type of {\it topological
    algebras}, singled out by the previous Theorem 4.1, are thus,
    in principle, {\it unital} ones, with a {\it compact
    spectrum}. Now, the first issue (:\;{\it ``unital''}) is
    actually referred to a technical item, pertaining to the {\it
    non-triviality of the spectrum of} $E$, the inductive limit
    topological algebra considered, by the
    aforesaid theorem (cf. (4.5)), as well as, to the {\it validity of} (4.7), both
    conditions being already pointed out in A.\;Mallios [11: pp.
    155, 156; (3.24), (3.29), along with Theorem 3.1 therein]. On
    the other hand, the hypothesis of {\it compactness of the
    spectra} involved can be relaxed to {\it local compactness},
    however, by still assuming {\it proper} ``transition maps'' in
    (4.7), while considering (sheaf) {\it cohomology with compact
    supports} (see G.E.\;Bredon, loc. cit.).
\end{minipage}

\bigskip
As already mentioned in the preceding, the previous type of
topological algebras is of a particular importance for a vast
extension of a differential-geometric character to spaces which
are not (smooth, viz. $\cc^\infty$-) manifolds, in the classical
sense of the term. Notice that $\cc^\infty (X)$, with $X$ a {\it
paracompact} (Hausdorff) $\cc^\infty$-{\it manifold}, is a {\it
topological algebra} of the previous type, whose {\it spectrum} is
(homeomorphic to) $X$, the same algebra being thus a {\it
geometric} one, with respect to $\cc^{\infty}_{X}$, the {\it sheaf
of germs of $\bc$-valued smooth} (:\;$\cc^\infty$-) {\it functions
on} $X$, where  the latter is a {\it fine sheaf} on $X$; see
A.\;Mallios [14: Chapts. III, XI]. Potential applications of the
aforesaid generalized perspective of the classical differential
geometry to problems, pertaining, for instance, to {\it quantum
gravity}, have been discussed, recently, in [19], [20], [21]; in
this connection, see also A.\;Mallios [15], [17], [18], as well
as, [22], [23].

\bigskip $ $\hfill
\begin{minipage}{11cm}
    \hspace{0.5cm}{\bf Scholium 4.1.}$-$ Motivated by the
    situation, that is relevant to the above Theorem 4.1, one is led to
    single out the following general point of view: Thus, by an
    {\it algebra space}, one means a pair,
    \[          
        (X,\ca ),
        \leqno{(4.13)}
    \]
    consisting of a {\it topological space} $X$ and a ($\bc$-){\it
    algebra sheaf} $\ca$ on $X$. In particular, we speak of a {\it
    geometric algebra space}, whenever (4.13), as before,
    satisfies, in addition, the corresponding herewith  \linebreak
\end{minipage}

$ $\hfill
\begin{minipage}{11cm}
    two
    conditions (4.2) and (4.3), as above. Yet, we say that (4.13)
    is a {\it compact algebra space}, if the topological space
    $X$, appeared therein, is, in particular, a {\it compact}
    ({\it Hausdorff}) space.
\end{minipage}

\bigskip \noindent
In this connection, what we have actually inferred, by the
previous Theorem 4.1, is simply that:

\bigskip \noindent (4.14) \hfill
\begin{minipage}{11cm}
    the {\it inductive limit of compact geometric algebra spaces}
    is a space of the {\it same sort}.
\end{minipage}

\bigskip \noindent
Indeed, the assertion is a straightforward application of the
proof of the aforesaid result (:\;{\it ``continuity theorem''} in
sheaf cohomology), according to the following data, with an
obvious meaning of the notation applied;
\[              
    \varinjlim (\ce_\al ,X_\al ):= (\ce \equiv \varinjlim \ce_\al
    , \,X\equiv \varprojlim X_\al ).
    \leqno{(4.15)}
\]
On the other hand, if the previous data are further supplied with
the appropriate {\it ``differentials''}, in the sense of {\it
``abstract differential geometry''} (ADG), cf. [14], one still
concludes that;

\bigskip \noindent (4.16) \hfill
\begin{minipage}{11cm}
    the {\it inductive limit of compact differential-geometric
    algebra spaces} is a space of the {\it same type}.
\end{minipage}

\bigskip \noindent
Now, both of the previous two propositions seems to stand in close
connection with several recent aspects in {\it quantum
relativity}; cf., for instance, [20], [18], [16].

\end{document}